\documentclass[a4paper]{IOP}

\usepackage[utf8]{inputenc}
\usepackage{comment}
\usepackage{graphicx}
\usepackage{amssymb}
\usepackage{amsmath}
\usepackage{blindtext}
\usepackage[bookmarks]{hyperref}
\usepackage{bookmark} 
\usepackage{caption}
\usepackage{subcaption}
\usepackage[makeroom]{cancel}
\usepackage[usenames, dvipsnames]{color}
\usepackage{array}
\usepackage{multirow}
\usepackage[english]{babel}
\usepackage[toc,page]{appendix}
\usepackage[acronym,toc]{glossaries}
\usepackage{fancyhdr}
\usepackage{wrapfig}
\usepackage{enumitem}
\usepackage[square,numbers]{natbib}
\usepackage{doi}
\usepackage{mathtools}
\usepackage{url}
\usepackage{fixmath}
\usepackage{authblk}
\usepackage{xcolor}
\usepackage{setspace}
\usepackage{footnotebackref}
\usepackage{titling}
\usepackage{soul}
\usepackage{acronym}
\usepackage{footnotebackref} 

\usepackage{amsmath} 
\usepackage[strings]{underscore}

\begin{document}

\pagestyle{fancy}
\fancyhf{}
\fancyfoot[CE,CO]{\thepage}
\fancyfoot[LE,RO]{}
\renewcommand{\headrulewidth}{0pt}


\title{Long Short-Term Memory for Early Warning Detection of Gravitational Waves}
\vspace{20pt}
\address{\large \bfseries Reem Alfaidi$^{1,2}$ and Christopher Messenger$^{1}$}
\vspace{10pt}
\address{$^{1}$ School of Physics and Astronomy, University of Glasgow, Glasgow G12 8QQ, UK}
\address{$^{2}$ The Department of Physics ,Tiba University, Madinah, Saudi Arabia}
\vspace{10pt}
\ead{r.alfaidi.1@research.gla.ac.uk}

\begin{abstract}
The pre-merger detection of gravitational waves from the early inspiral phase of compact binary coalescence events would allow the observation of the earlier stages of the merger in the electromagnetic band. This would significantly impact multi-messenger astronomy, giving astronomers potential access to rich new information. Here, we introduce a proof-of-concept deep-learning-based approach to produce pre-merger early-warning alerts for binary black hole systems. We show the possibility of using a Long Short-Term Memory network trained on the whitened detector strain in the time domain to detect and classify compact binary events. In this work, we consider a single advanced Laser Interferometer Gravitational-Wave Observatory detector at design sensitivity and make approximate sensitivity and early warning capability comparisons with approximations to traditional matched filtering approaches. We find that our model is competitive in both aspects, and when applied to a simulated test dataset was able to produce an early alert up to four seconds before the merger.
\end{abstract}

\acrodef{BCE}[BCE]{binary cross entropy}
\acrodef{BBH}[BBH]{Binary Black Hole}
\acrodef{BNS}[BNS]{Binary Neutron Star}
\acrodef{CBC}[CBC]{Compact Binary Coalescence}
\acrodef{CLDNN}[CLDNN]{Convolutional, Long Short-Term Memory, Fully-Connected Deep Neural Network Model}
\acrodef{CNN}[CNN]{Convolution Neural Network}
\acrodef{CSNR}[CSNR]{Cumulative Signal-to-Noise Ratio}
\acrodef{EM}[EM]{Electromagnetic}
\acrodef{GRB}[GRB]{Gamma-Ray Burst}
\acrodef{GW}[GW]{Gravitational Wave}
\acrodef{GWTC-3}[GWTC-3]{Third Gravitational-Wave Transient Catalogue}
\acrodef{LIGO}[LIGO]{Laser Interferometer Gravitational-Wave Observatory}
\acrodef{LSTM}[LSTM]{Long Short-Term Memory}
\acrodef{NSBH}[NSBH]{Neutron Star–Black Hole}
\acrodef{PSD}[PSD]{Power Spectral Density}
\acrodef{RNN}[RNN]{Recurrent Neural Network}
\acrodef{ROC}[ROC]{Receiver Operating Characteristic}
\acrodef{SNR}[SNR]{Signal-to-Noise Ratio}

\section{Introduction} \label{sec: 1}

%
A new era in the study of the universe began with the detection of \acp{GW} from Compact binary coalescence \cite{Abbott16}. In the years after the first direct detection of \ac{GW}s from a \ac{BBH} merger, the field of gravitational-wave astronomy has grown dramatically. Since then, numerous events have been reported by \ac{LIGO}-Virgo \cite{Advanced,Acernese}, many of which were detected in low-latency $\mathcal{O}(\text{mins})$. Detecting \ac{GW}s from compact binaries has become routine, particularly from \ac{BBH}s signals, nonetheless, \ac{GW}s from \ac{BNS} and \ac{NSBH} mergers are still relatively rare. These type of events are of particular interest given their potential for counterpart \ac{EM} signals~\cite{Abbott21,Abbott2021}. During the third observing run (O3) ~\cite{Abbott2020}, which was split into two segments (O3a and O3b), over 70 additional \ac{GW} events were detected. Among these is the discovery of a second \ac{BNS} merger. With the addition of the O3 events, the \ac{GWTC-3} has been released ~\cite{Abbott23}, and contains more than 90 events including examples of all possible configurations of compact binary object mergers. 

In 2017, Advanced \ac{LIGO} and Virgo detected \ac{GW}s signal from the merging of two neutron stars~\cite{Abbott1717}. At 1.7 seconds after the merger of these two neutron stars, the Fermi \ac{GRB} Monitor observed a \ac{GRB}~\cite{Abbott17}. The association between the two observations led to an \ac{EM} follow-up within the sky localisation area of \ac{GW}s and initiated a new era in multi-messenger astronomy~\cite{Abbott77}. Multi-messenger astrophysics aims to use information about the astrophysical universe provided by all four fundamental forces of nature: gravity, weak forces, strong forces, and \ac{EM} forces, the latter of which provide unique information about their sources~\cite{Peter}. However, the most daunting challenge of obtaining prompt \ac{EM} observations is the rapid detection and identification of \ac{GW}s. If we can quickly locate the signals of \ac{GW}s, we can promptly initiate \ac{EM} observations \cite{Maricab}.

Matched-filtering, a cornerstone data analysis technique employed by \ac{LIGO} and Virgo, plays a critical role in detecting compact binary \ac{GW} signals. This method has successfully identified \ac{BBH} coalescences\cite{Leo,Sushant,Yelikar,Mukesh,Ryan,Alexander20}. It involves a detailed analysis of observational data using a comprehensive library of pre-constructed theoretical waveform templates. By continuously examining data streams, matched-filtering generates candidate signals, often called triggers. While this technique is instrumental in searching for \ac{GW}s, it has challenges. One notable limitation is the potential reduction in data processing speed, particularly when the analysis involves a substantial template bank. This factor necessitates carefully balancing the template library's comprehensiveness and the data processing efficiency.

Deep learning, a subset of machine learning, has grown increasingly popular due to the rapid advancement of the graphics processing unit technology \cite{Laith}. This approach has succeeded in various fields, including medical diagnosis, gene expression classification, and image processing \cite{Deepak,Rajit,Shervin}. In astrophysics, deep learning has become an indispensable technique for analysis and has significantly advanced gravitational-wave \ac{GW}s detection \cite{Timothy,Hunter,Daniel,Marlin}. Deep learning has been particularly beneficial for analyzing \ac{GW} data in low latency and identifying \ac{GW} event signals~\cite{Gregory}. It has also shown promise in signal identification and glitch categorization within \ac{GW}s astronomy \cite{Zevin}, where deep learning was initially demonstrated as an effective detection method. As the most computationally demanding stages are precomputed during training, deep learning enables rapid analyses, leading to low-latency searches that are potentially much faster than other similar categorization techniques.

An early application of a supervised machine learning algorithm~\cite{Paul} was proposed as an alternative to the traditional chi-squared statistic used in post-processing matched-filter output. Since then, several groups~\cite{Elena} have developed other interesting ways to perform \ac{CBC} signal recognition using deep learning-based methodologies. Several studies have used \ac{CNN} algorithms to look for \ac{GW}s from \ac{CBC} signals. This research \cite{Kyungmin} discusses whether deep learning can be applied to \ac{CBC} \ac{GW} searches. An \ac{RNN} algorithm, Bayesian Neural Networks, was developed to detect \ac{BBH} signal~\cite{Yu}. It can identify the entire simulated duration (8~sec) of the \ac{GW} event, including the inspiral stage. The \ac{CLDNN} model, a combination of \ac{CNN} and \ac{LSTM}, is incorporated with the Bayesian approach. This model claims to detect all \ac{BBH} events in the \ac{LIGO} Livingston O2 data, and also claims detection of 90\% of simulated events with SNR $>7$\footnote{It is not clear at what false alarm threshold these sensitivities are obtained.}. 

In this paper, we aim to use a \ac{RNN} known as an \ac{LSTM} to detect the signal of \ac{GW}s and focus on the early warning detection of the signal before the merger time of the \ac{CBC}. This is a single detector analysis and we present a preliminary test case that uses \ac{BBH}s mergers as a foundational dataset for developing our early warning detection algorithm. It is important to note that while the primary objective of the algorithm extends beyond \ac{BBH}s, these events serve as an ideal starting point due to their characteristic short duration, which allows for developing in-depth analysis without the challenge of managing massive data sets and well-defined signal patterns. In this context, \ac{BBH} mergers play a key role in improving the sensitivity and fine-tuning of the model's parameters. This lays a solid basis for future modifications that seek to detect more complicated and diverse signals, like those with lower mass and correspondingly longer time-series, numerous input channels acting as detectors, and potentially spinning systems. While the current test case demonstrates the algorithm's capability for early warning detection of signal mergers, it is crucial to acknowledge that it does not currently offer early sky localization estimates. Such estimates are vital for the practical application of the algorithm in real-world scenarios, particularly in multi-messenger astronomy, where prompt localization is essential for follow-up observations. Recognizing the importance of this feature, we defer the challenge of integrating rapid and early warning sky localization to future research. The paper is organised as follows. In Sec. \ref{sec: 2}, we introduce the basic ideas of \ac{LSTM} networks and discuss the type of \ac{LSTM} we used. In Sec. \ref{sec: 3}, we describe the data generation and the training details of the \ac{LSTM}. In Sec. \ref{sec: 4}, we cover the design and development of the \ac{LSTM} model. We present the results of the \ac{LSTM} model in Sec. \ref{sec: 5}. Finally, we conclude our work in Sec. \ref{sec: 6}.

\section{Recurrent Neural Networks} \label{sec: 2} 

An \ac{RNN} is a particular type of artificial neural network adapted to deal with the sequential data within the scope of deep learning~\cite{Zachary}. This distinctive architecture empowers \acp{RNN} to grasp and leverage information from preceding time steps, rendering them particularly effective in analyzing time series data. This type of network has achieved excellent performance with sequential data over the past few years. For example, \ac{RNN} applications are used in Apple's Siri and Google's voice search~\cite{Tianyu}. In the context of \ac{GW}s detection, \ac{RNN}s offer significant advantages. By training on a large dataset of labeled \ac{GW}s signals, \ac{RNN}s can learn to recognize and classify different waveforms. This capability is crucial for identifying and distinguishing \ac{GW}s signals from background noise. They also benefit from a reduced relative size and complexity of the network in comparison to other deep learning approaches, such as \acp{CNN}, where the input layer is receptive to the entire signal length. The purposeful retention of relevant information in an \ac{RNN} allows it to focus on short sections of the input time-series sequentially. 

\begin{figure}
\includegraphics[width= 1 \linewidth]{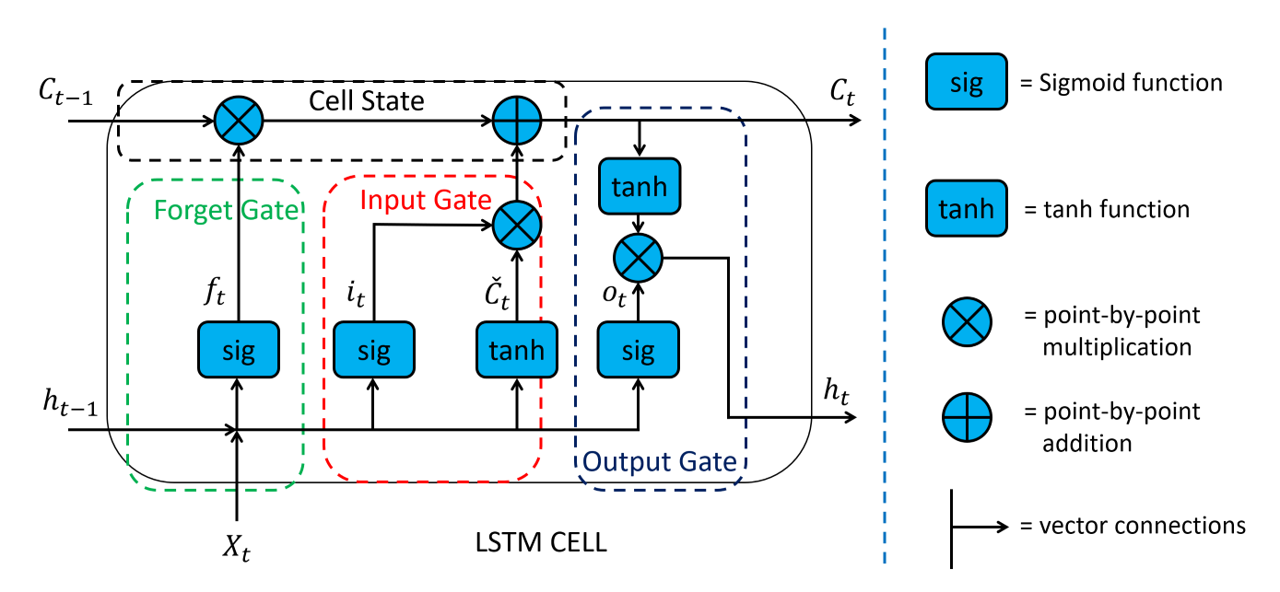}
\centering
\caption{Simple structure of \ac{LSTM}~\cite{Smagulova}, where $x_{t}$ is a current input, and $h_{t-1}$ is the output of the last \ac{LSTM}  unit and $c_{t-1}$ is the memory of the last \ac{LSTM} unit, $c_{t}$ is the updated memory, and $h_{t}$ is the current output. The sigmoid and tanh functions are the activation functions. \label{fig: 1}}
\end{figure}

An \ac{LSTM} network is a class of \ac{RNN}, which was first presented in 1997~\cite{Zachary}. It is designed to solve many tasks previously unsolvable using \acp{RNN}~\cite{Yong}. Moreover, using \acp{LSTM} can avoid gradient descent problems~\cite{Sepp} found in \acp{RNN}. It has achieved credible results in various applications, such as language modeling, speech-to-text transcription, and machine translation~\cite{Alex}. The fundamental concept of an \ac{LSTM} is built around a cell state that functions as long-term memory. The cell state acts as a transport highway that transfers relative information step by step until the end of the sequence chain. It consists of three gates: input gate, forget gate, and output gate, as shown in Fig.~\ref{fig: 1}, 
where the gates control the information that goes through the \ac{LSTM}. The forget gate decides which data should be discarded from the cell. The input gate decides which values from the current input should be used to update the memory cell. Finally, the output gate decides which output will be presented based on the input and the memory of the cell. The following set of equations can illustrate operations in an \ac{LSTM} layer as follows:
\begin{equation} \label{eq: 1}
i_{t} = \sigma \left( w_{i} \cdot [\left[h_{t-1},x_{t} \right] + b_{i} \right),
\end{equation}
\begin{equation} \label{eq: 2}
f_{t} = \sigma \left( w_{f} \cdot \left[h_{t-1},x_{t} \right] + b_{f} \right),
\end{equation}
\begin{equation} \label{eq: 3}
o_{t} = \sigma \left( w_{o} \cdot \left[h_{t-1},x_{t} \right] + b_{o} \right),
\end{equation}
where $i_{t}$ is the input gate, $f_{t}$ is the forget gate, $o_{t}$ is the output gate, $\sigma$ is the sigmoid function, and $w$ is set of weights associated with each gate performing a linear transformation represented by a matrix. For the respective gate neurons ($x$), $h_{t-1}$ is the output of the previous \ac{LSTM} block at timestamp $t-1$, $x_{t}$ is the input at the current timestamp, and $b_{x}$ represents the biases for the respective gates ($x$). A bias is applied as an intercept in a linear equation; it is a parameter in the neural network, together with the network weights, that are used to change the output. The following set of equations then describe the cell state of the \ac{LSTM} layer:
\begin{equation} \label{eq: 4}
\Tilde{c}_{t} = \tanh{\left(w_{c} \cdot \left[h_{t-1},x_{t} \right] + b_{c}\right)},
\end{equation}
\begin{equation} \label{eq: 5}
c_{t} = f_{t} * c_{t-1} + i_{t} * \tilde{c}_{t},
\end{equation}
\begin{equation} \label{eq: 6}
h_{t} = o_{t} * \tanh{\left(c_{t}\right)},
\end{equation}
where $*$ represents the element wise multiplication of the vectors, $c_{t}$ is cell state (memory) at timestamp $t$, and $\Tilde{c}_{t}$ is the potential update to the cell state at timestamp $t$. The cell state gives the \ac{LSTM} power to learn long strings of sequential data successfully~\cite{Shrestha}. However, various \ac{LSTM} models have different architectures depending on the number of hidden layers.

%
In a stacked \ac{LSTM} architecture (as used in the analysis presented in this paper), the arrangement is such that several \ac{LSTM} layers are stacked on top of each other, creating a deeper network structure. In this configuration, each \ac{LSTM} layer of the stack processes the input sequence and then forwards its output to the subsequent layer. This layered approach enables the network to discern and learn intricate features at varying degrees of abstraction. Essentially, the output of one \ac{LSTM} layer becomes the input for the next, allowing for a more nuanced understanding and processing of the sequential data.

\section{Data Generation Process} \label{sec: 3}
In this section, we will discuss the need for simulating \ac{GW} signals. Our \ac{LSTM} model requires thorough training, validation, and testing to ensure its effectiveness. To achieve this, we must have access to numerous \ac{GW} signals across various noise realizations. This process of simulating \ac{GW} signals is essential for robustly evaluating and fine-tuning our model's performance.

Simulated \ac{GW}s are generated using the \href{https://pycbc.org/}{PYCBC} library. Our study concentrates explicitly on \ac{BBH} signals, as opposed to incorporating \ac{BNS} systems, due to the inherently shorter length of higher mass \ac{BBH} signals. 

\begin{table}
\centering
\caption{The prior distribution used on the \ac{BBH} signal parameters.}
\label{tab: 1}
\begin{tabular}{cccc}
\hline \hline
Parameter  & distribution & Min. & Max. \\ 
\hline  
Mass 1 $(m_{1})$  & $\propto m_{1}^{?}$ & $5 M_{\bigodot}$  & $95 M_{\bigodot}$ \\
Mass 2 $(m_{2})$  & Uniform & $5 M_{\bigodot}$  & $m_1$ \\
Phase at Coalescence $(\phi_{\circ})$  & Uniform & $0$  & $2\pi$ rad. \\
Right Ascension $(\alpha)$  & Uniform & $0$       & $2\pi$ rad. \\
Sine Declination $(\sin(\delta))$  & Uniform & -1 & 1 \\
Cosine Inclination $(\cos(\Theta_{jn}))$  & Uniform & -1       & 1 \\
Polarisation $(\psi)$        & Uniform & $0$        & $\pi$ rad. \\
Coalescence time: Training            & Uniform & $4$       & $9$ seconds.\\
Coalescence time: Testing             & Uniform & $8$       & $13$ seconds.\\
Optimal \ac{SNR} $(\rho_{\text{opt}})$: Training  & Uniform & $7$ & $20$ \\
Optimal \ac{SNR} $(\rho_{\text{opt}})$: Testing  & Uniform & $2$ & $20$ \\
\hline
\end{tabular}
\end{table}

\begin{figure}
\centering
{\includegraphics[width=0.6\textwidth]{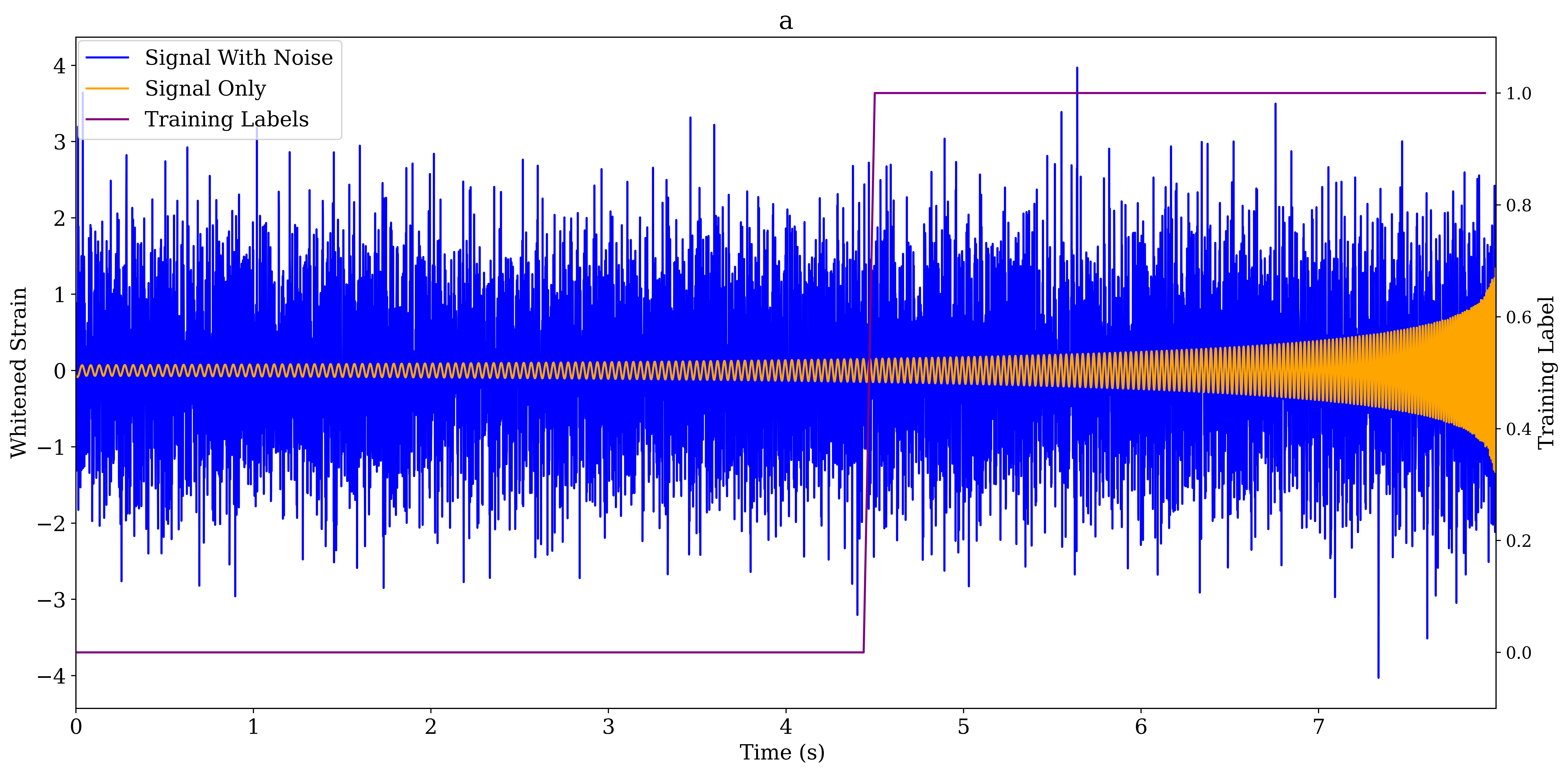}}
{\includegraphics[width=0.6\textwidth]{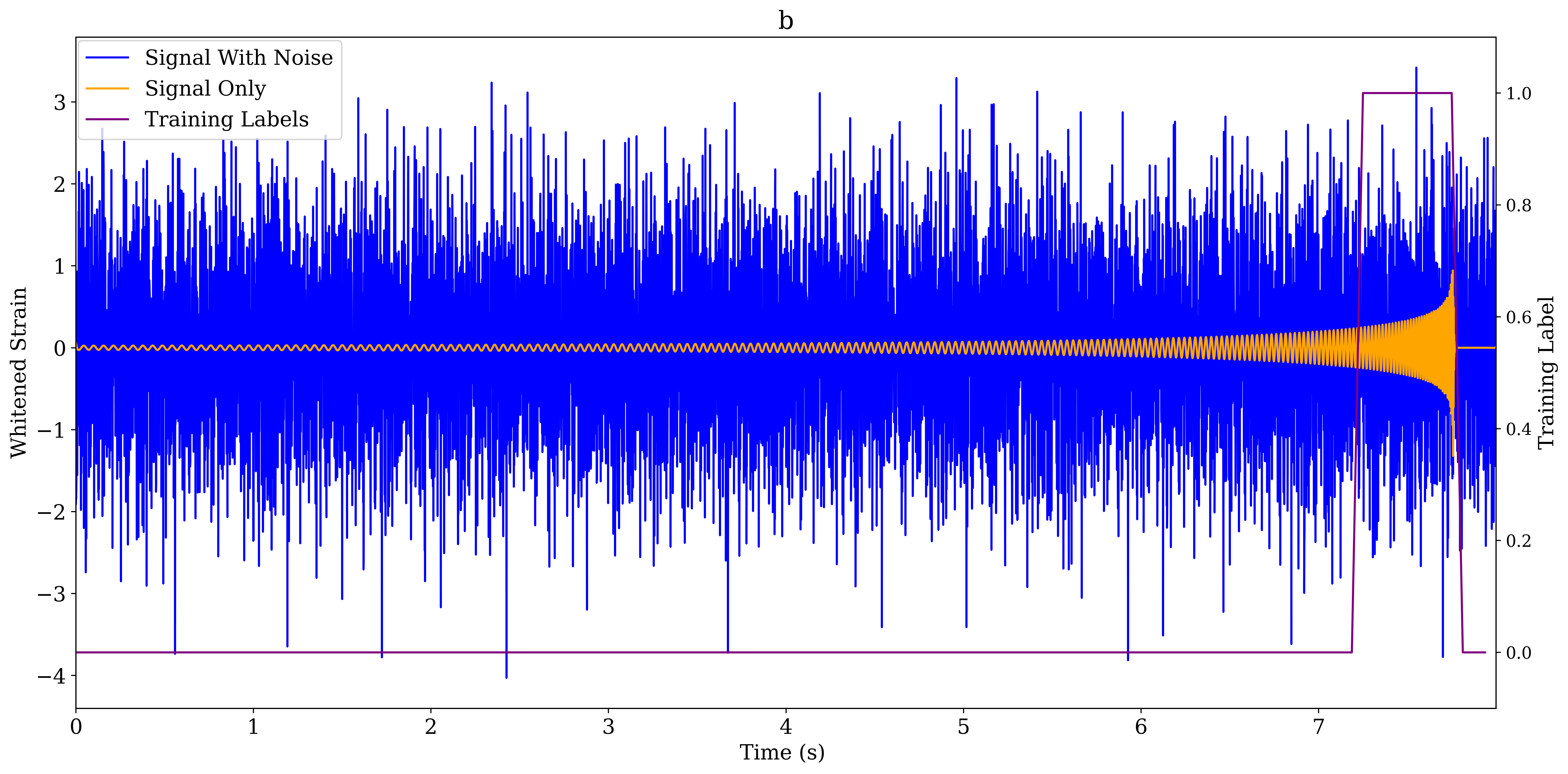}}
{\includegraphics[width=0.6\textwidth]{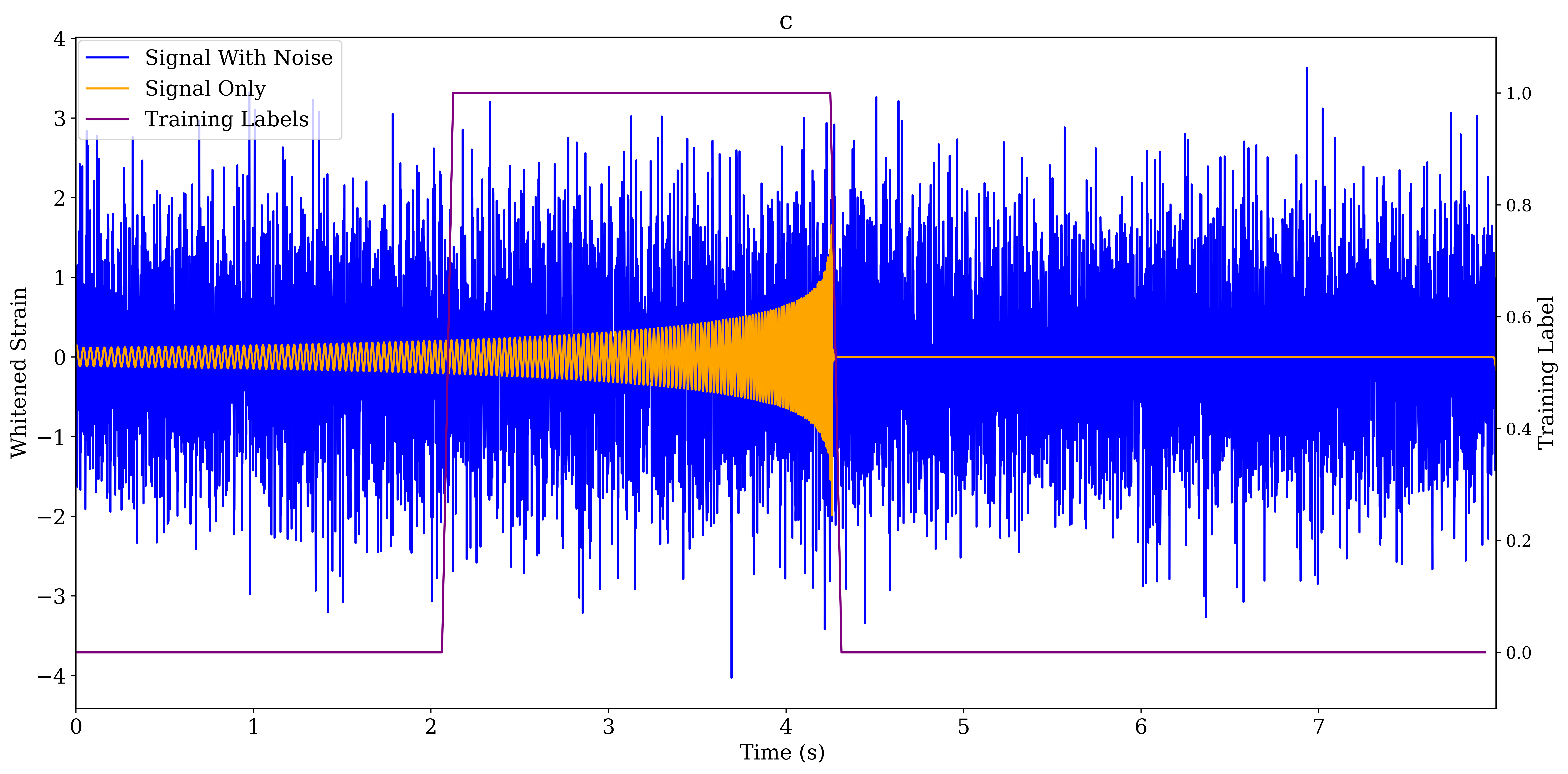}} 
{\includegraphics[width=0.6\textwidth]{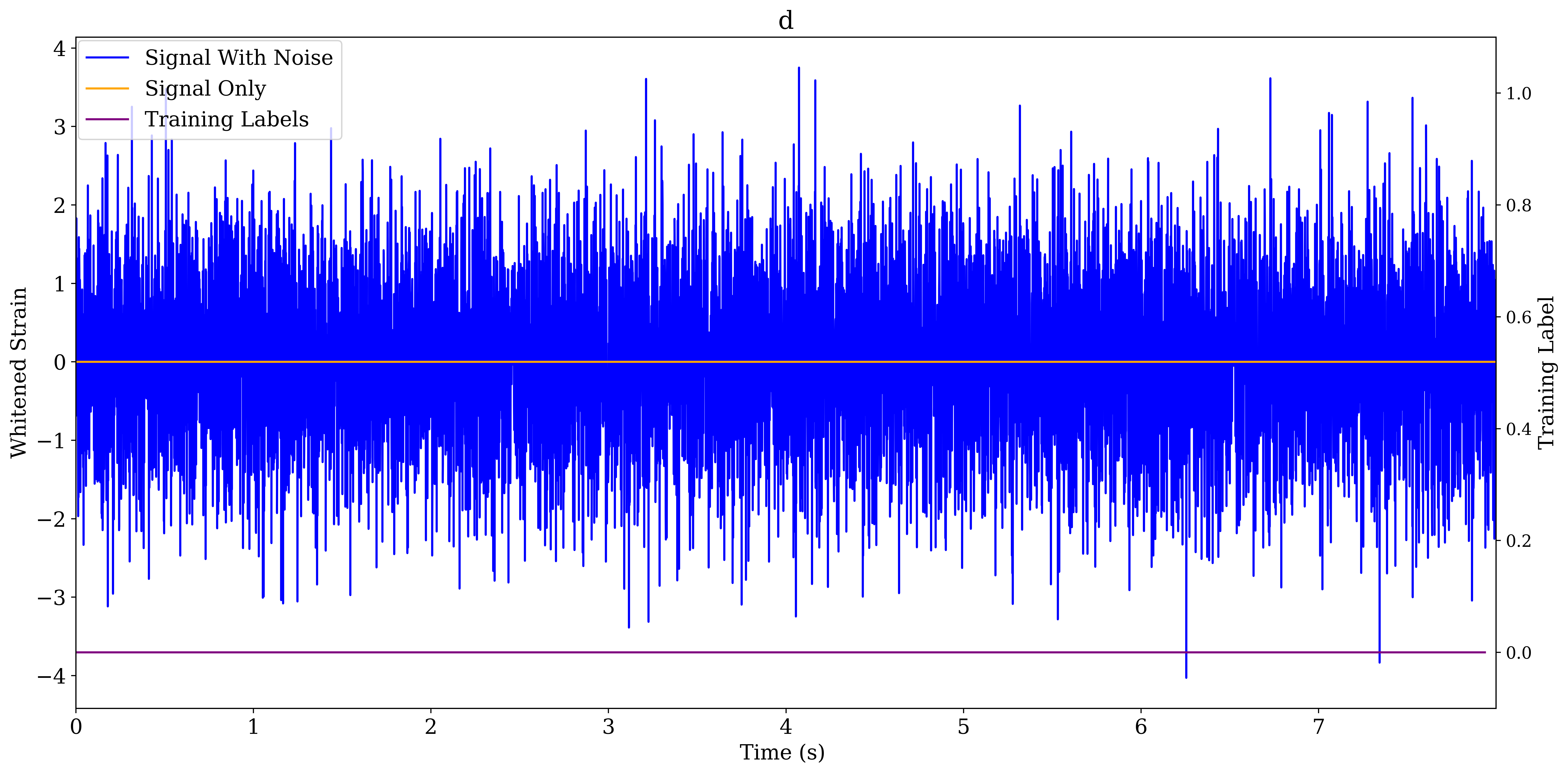}}   
\caption{Example time series of simulated data used for training the \ac{LSTM} model, showcasing various combinations of signal and noise. Each plot captures a unique scenario with signals merging in different regions of the time-series and an example featuring a noise-only case. (a) Signal merger outside the training window, \ac{SNR} $\rho_{\text{opt}}$ = 17.85, chirp mass = 6.95 $M_{\odot}$. (b) Signal merger just prior to the end of the training window, \ac{SNR} $\rho_{\text{opt}}$ = 8.3, chirp mass = 8.97 $M_{\odot}$. (c) Signal merger early in the prior window, \ac{SNR} $\rho_{\text{opt}}$ = 16.27, chirp mass = 7.24 $M_{\odot}$. (d) Noise-only scenario.}
\label{fig: 2}
\end{figure}

We randomize the parameters according to a distribution to simulate the \ac{GW} waveform as listed in Tab. \ref{tab: 1}. We generate several samples to model different astronomical parameters. Each parameter is constructed to match a specified size, ensuring a comprehensive and randomized sampling across these parameters. This approach allows for a robust and varied dataset, reflecting a realistic distribution of signal parameters. specifically the H1 \ac{LIGO} Hanford detector. Specifically, our analysis focuses on systems with mass ranges from $5M_{\odot}$ to $95M_{\odot}$. We employ a power-law mass distribution for selecting $m_1$, ensuring $m_1 > m_2$. For $m_2$, the masses are uniformly chosen between 5 $M_{\odot}$ and the value of $m_1$. The sampling is conducted at 1024 Hz, with the spin parameter set to zero. The data simulated is based on the signal from a single detector. The optimal \ac{SNR} of the simulated signals is given by
\begin{equation} \label{eq: 7}
\rho_{\text{opt}}^{2} = \frac{4}{d^{2}}\displaystyle\int_{0}^{\infty} \dfrac{| \tilde{h}_{d=1}(f) |^{2}}{S_{n}(f)} \; df,
\end{equation}
where $\tilde{h}_{d=1}(f)$ is the Fourier transform of the signal computed at a distance of $1$ Mpc, and $S_{n}(f)$ is the \ac{PSD}. We control the \ac{SNR} distribution by assigning the distance $d$ and correspondingly scaling the signal amplitude based on the desired \ac{SNR}. We have chosen to use an \ac{SNR} distribution that is uniform in the range between 7 and 20 for training the model, and for testing the model we use a range between 2 to 20. 
We generate signals using the SEOBNRv4 opt approximant~\cite{Sascha}. To simulate the signal from the SEOBNRv4 opt waveform, we initially employ different sampling frequencies depending on the system's mass. For systems where the mass is less than 25 $M_{\odot}$, a sampling frequency of 4096 Hz is used. Conversely, for systems with a mass greater than 25 $M_{\odot}$, the sampling frequency is set to 2048 Hz. Following the initial simulation, we resampled the signal to the sampling frequency of 1024 Hz. This resampling process standardizes the data for subsequent analysis and ensures consistency across different mass ranges but will suppress high frequency information (close to merger) for the lower mass systems. We compute the signal's plus and cross-polarisation and from this construct $h(t)$ by applying the detector antenna pattern, a function of the randomised sky location and polarisation angles.
We generate Gaussian noise using a specified \ac{PSD}, aLIGOZeroDetHighPower~\cite{LIGODocument}. Then, we add the signal to the noise; examples of a training data time-series can be seen in Fig.~\ref{fig: 2}. These diverse data samples are critical for effectively training the model to distinguish between noise and meaningful patterns. This approach enhances the model's ability to identify signals amidst noise and improves its effectiveness as an early warning system by learning to detect signals that merge at unpredictable times. We also performed a matched-filtering analysis on the dataset, which will be used later to compare with the results obtained from the \ac{LSTM} model. We whitened the data before using it in our \ac{LSTM} algorithms. We generated a waveform with 12 seconds with different merger times, where the merger was located between 4 to 9 seconds. Then we took the first 8-second window where the merger time can be inside or outside these 8 seconds for training the model while testing the model in longer data, 12 seconds, and the merger is located between 8 and 13 seconds.

For training (and testing) the \ac{LSTM} model we must provide labels indicating the presence (1) or absence (0) of a signal at each step as the model processes the times-series. We construct these labels based on the optimal cumulative \ac{SNR} of times-series containing signals and the merger time of those signals. The cumulative \ac{SNR} is defined as
\begin{equation} \label{eq:csnr}
\rho_{\text{C}}^{2}(t) = \left\{ \begin{array}{lr}
    2 \int\limits_{0}^{t} |h_{\text{w}}(t)|^2\,dt&\text{if}\,t\leq t_{c} \\
    0 &\text{otherwise}
\end{array}\right.
\end{equation}
where $h_{\text{w}}(t)$ is the whitened signal time-series, and the cumulative \ac{SNR} is reset to zero following the merger at $t=t_{c}$. Time steps are classified as not containing a signal if the cumulative \ac{SNR} $<5$ and classed as containing a signal if $\geq 5$. For time-series not containing any signal the optimal cumulative \ac{SNR} is zero and all time steps are labelled as the noise-only class (0).  

As the \ac{LSTM} gives a prediction every 64 time steps (a prediction rate of 16 Hz), we have 128 labels for each time-series of the training dataset and 192 labels for each time series of the testing dataset. Additionally, $25\%$ of our data does not contain any signal, which is the noise class only, and the rest of the data are signal plus noise. Data are first whitened and then normalized using the Z-score. The training set consists of 1 million \ac{GW}s events, which include either signal with noise or noise only. The validation set comprises 100,000 times-series, and the test set contains 20,000 times-series.

\section{The LSTM Model} \label{sec: 4}

In this section, we explore the design and implementation of a four-layer stacked \ac{LSTM} model. We discuss the architecture and the training process, including optimization strategies. We then examine the testing phase where we evaluate the model under different scenarios. Finally, we discuss how to compare the performance of our \ac{LSTM} model with traditional matched filtering techniques.

\subsection{The LSTM Design} \label{subsec: 4.1}
Using the PyTorch~\cite{PyTorch} library, we adopt a stacked \ac{LSTM} configuration comprising four layers, each using 32 neurons. The input dimension is 64, meaning that 64 time-series samples are read as input to the network at each step. The final \ac{LSTM} output is passed to a dense layer with 16 neurons and then an output layer of dimension 2. As we deal with a classification problem, we use the \ac{BCE} with logits loss as the loss function. The \ac{BCE} with logits loss function measures an effective average difference between
the predicted probabilities (after applying a sigmoid function to the network outputs) and the true labels in a binary classification scenario. Hence the output statistics from the network are unbound but are converted to probabilities in the range $[0,1]$ within the loss function. Prior to our analysis the data were categorised into training, validation and test data. The training and validation data are generated from the same distribution and hence, 10\% of the training data described in Sec.~\ref{sec: 3} is allocated as validation data.

When in operation, the model will iteratively take in input in blocks of 64 time-series samples (equivalent to 16 steps over 1 second of data). Together with this input, the \ac{LSTM} will combine information passed to it from the previous analysis step in the form of the cell state memory and process this input through the \ac{LSTM} network to obtain a prediction for the presence or absence of a signal at that step. It will also update its memory state for passing on to the subsequent step. Therefore, the primary output from the model has the form of a lower sample rate (16 Hz) time series of predictions.
In building our model for sequence data analysis, we strategically chose four stacked LSTM layers to adeptly handle long-term data dependencies, enhancing its pattern recognition capability. Opting for the sigmoid function as the output layer aids in binary classification, providing probabilities for data categorization. We initiated the model with 32 neurons in the input layer, striking a balance between capturing complex patterns and avoiding overfitting, ensuring both computational efficiency and effectiveness in processing intricate temporal data. These decisions reflect our careful consideration of the model's complexity and operational feasibility.

\subsection{Training the LSTM Model} \label{subsec: 4.2}

\begin{figure}
\includegraphics[width=\textwidth]{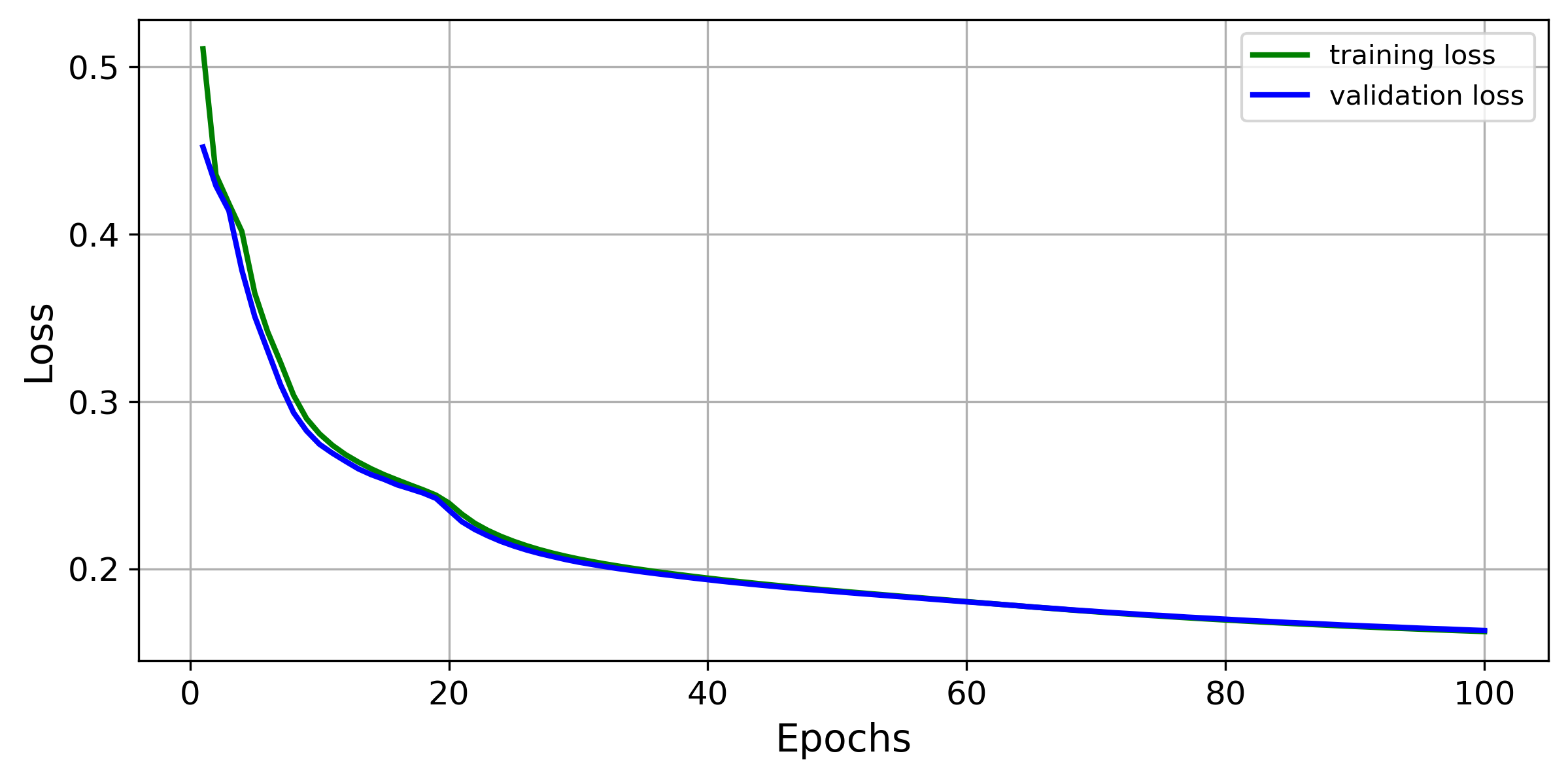}
\centering
\caption{The evolution of the training and validation \ac{BCE} loss during training as a function of epoch. This trend indicates a consistent improvement in the model's performance as the epochs progress.\label{fig: 3}}
\end{figure}

The \ac{LSTM} model required 33 hours to train fully and was conducted on an NVIDIA GeForce RTX 3090 GPU equipped with 24GB of memory. The number of epochs used during the training is 100, with a batch size of 1000 and a learning rate of $1\times 10^{-4}$. In our model, a minimum of $7 \times 10^{5}$ training time series were required to prevent over-fitting, and to converge on our most sensitive results, $1$ million training time-series were used. The model's performance was observed by analyzing the loss curves for the training and validation datasets. The loss curves, shown in Fig.~\ref{fig: 3}, indicate effective training, where the training loss curve steadily declines, showing that the model could efficiently learn from the training set. Simultaneously, the validation loss curve descends in step with the training loss, with little deviation from it. This parallel fall shows that the model was not over-fitting to the training set and is a sign of its ability to generalize to unseen data. Training was stopped after 100 epochs (equivalent to $\sim 1.5$ days) although we note that the training had not yet converged. Further training would likely improve the model but those improvements would be minor given the final gradients of the loss curves.

The choices made regarding training hyper-parameters were deliberate; for instance, the batch size was selected to balance the computational load with the model's ability to generalize from the training data. The learning rate was chosen to ensure steady progress without overshooting the minima in the loss landscape. The training data volume was sufficient to represent the problem space without causing excessive training times or requiring prohibitive memory resources. We made these choices to build a strong and efficient training process suited to the needs and limits of our computational environment.

\subsection{Testing the LSTM Model} \label{subsec: 4.3}

To evaluate the efficacy of our model, we conducted a series of tests using an independent test dataset. These tests were designed to assess the model's performance using a number of metrics. The test dataset, distinct from the training and validation datasets, comprises 20,000 events, each extending over 12 seconds, and features a uniform distribution of \ac{SNR} ranging from 2 to 20. We used 20,000 time series to get enough data for reliable false alarm rate calculations. Each series has 192 labels, making a total of 3,840,000 data points. After excluding the first 4 seconds, we're left with 256,000 data points, including 2,116,459 noise and 443,541 signal points. This calculation is pivotal for our detection objectives, particularly in achieving false alarm rates down to one per day due to detector noise. Such a specific false alarm rate is essential for ensuring the reliability and precision of our model in real-world scenarios. Based on rate predictions for \ac{BBH}s~\cite{Abbott20}, where it is expected that in the current 4$^{\text{th}}$ advanced detector observing run observes $\mathcal{O}(100)$ such events, a false alarm rate of 1 per day could be considered tolerable given the expected comparable true alarm rate. In early warning analysis, we use the LSTM model to analyze test data, and it operates with small time steps (1/16 second) as described in Sec \ref{subsec: 4.1}. For each of these time steps, we calculate a statistic. If this statistic exceeds a specific threshold, which is determined based on the likelihood of experiencing a false alarm once per day, we classify it as an early warning trigger. However, it's essential to note that we only label it as a trigger if it's the first time it has surpassed this threshold within a 12-second long signal test data time-series.

\subsection{Early warning comparison with matched-filtering}\label{subsec: 4.4}

In order to provide an approximate benchmark to compare against our \ac{LSTM} early warning results, we have roughly approximated the sensitivity of a non-machine learning search method similar to matched filtering. We calculate the \ac{CSNR} using an exact template to provide early warnings, akin to how matched filtering would accumulate this statistic in real time. Our approach also broadly equates realistic false alarm thresholds with a set of specific \ac{CSNR} values. As a result, we explore various \ac{CSNR} thresholds to compare the \ac{LSTM} model results (which are defined at specific false alarm rates). We approximate the early warning time as the duration between the merger and when the \ac{CSNR} crosses these thresholds. This comparison aims to approximately evaluate the effectiveness of our approach in early signal detection scenarios.

\subsection{Detection sensitivity comparison with matched-filtering}\label{subsec: 4.5}

We also aim to approximately compare the raw sensitivity of the \ac{LSTM} analysis to existing matched-filter methods (irrespective of the early warning output). In this instance we do not use existing matched filtering algorithms for early warning detection, such as those discussed in the referenced literature. Instead, we opt for a simpler comparison. We compared the \ac{LSTM} results with a hypothetical scenario where the exact mass template is known without the need to search over a template bank. Here, the detection statistic is the matched-filter \ac{SNR}, computed using PyCBC and maximized over a 4-second window for the signal's arrival time. In our exact template matched-filter analysis, random templates from our training mass prior are generated when applying this single template approach to instances of noise only. To be consistent we also maximise over the same 4-second window when obtaining the corresponding statistic from the \ac{LSTM} analysis.

This approximate comparison sets an unrealistic upper-bound for sensitivity but it helps contextualize our results. As stated in the previous section, a similar method is used as a benchmark for the \ac{LSTM} early warning analysis, where we use a single true mass template for each test signal and compute the optimal \ac{CSNR} as a function of time. Our aim in both cases is to determine whether our proof-of-principle \ac{LSTM} analysis can get close to current levels of detection capability and early warning times. 

\section{Results} \label{sec: 5}

After completing the training and validation steps, we evaluate the performance of the \ac{LSTM} model over a testing dataset as described in section \ref{sec: 4}. Our primary focus has been on two key goals: firstly, approximating the sensitivity of \ac{LSTM} models in comparison to the matched filtering approach, and secondly, investigating the effectiveness of \ac{LSTM} models for detecting signals during the early stages before a merger event.

\subsection{LSTM Model detection sensitivity} \label{subsec: 5.1} 

\begin{figure}
\includegraphics[width= 0.8 \linewidth]{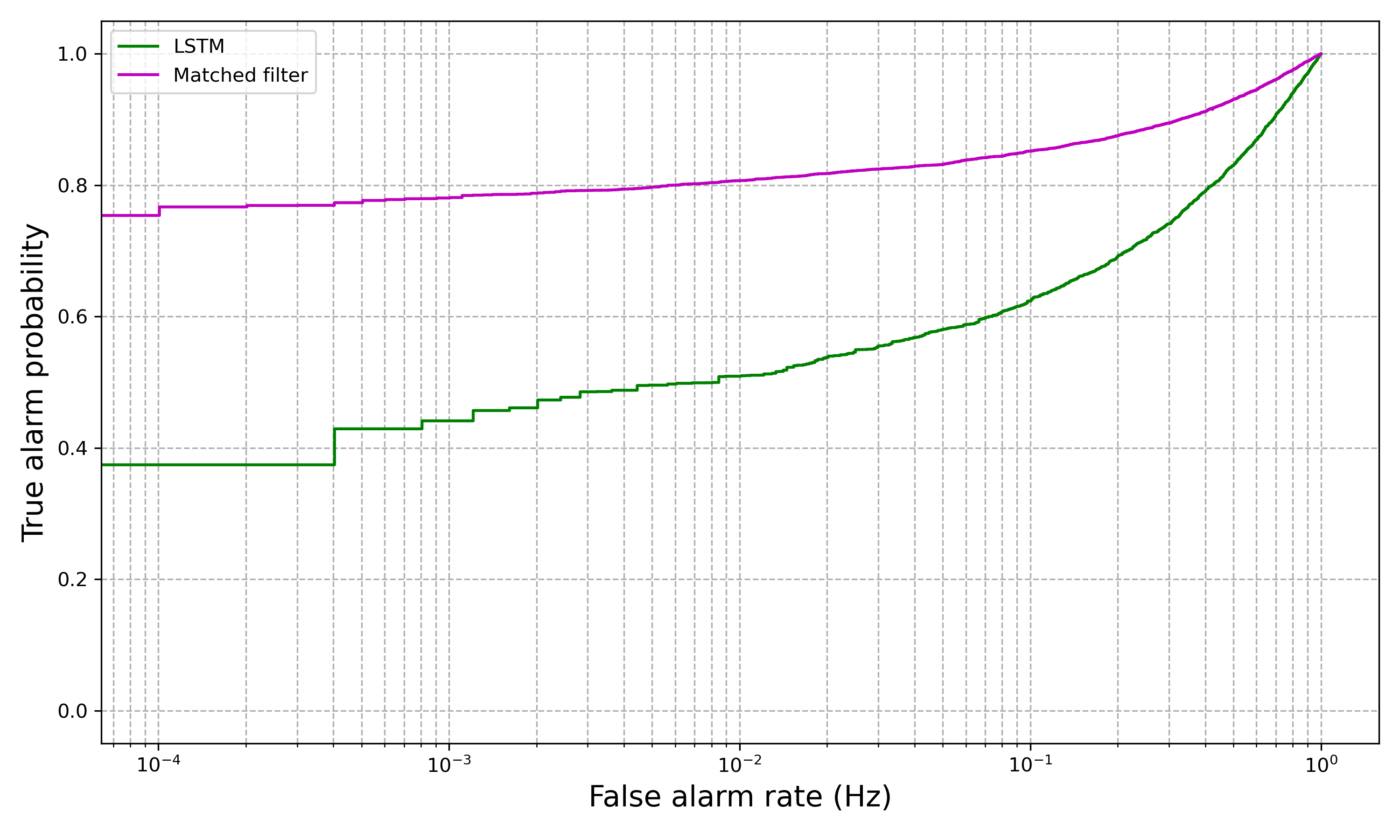}
\centering
\caption{\ac{ROC} curve comparing the detection sensitivity of an \ac{LSTM} model (green line) with that of a matched filtering approach using an exact template (magenta line) where both statistics have been maximised over a 4-second window. The \ac{ROC} curve demonstrates the true alarm probability as a function of the false alarm rate.\label{fig: 4}}
\end{figure}

In our analysis, we compared the \ac{LSTM} model to an approximate version of matched-filtering for detecting \ac{GW} signals. We used \ac{ROC} curves to evaluate how well the \ac{LSTM} model performs compared to matched filtering. \ac{ROC} curves help us visually compare how accurately each method identifies true positives as a function of false alarm thresholds. The \ac{ROC} curve in Fig.~\ref{fig: 4} shows how the \ac{LSTM} model and matched filtering compare in terms of sensitivity when applied to the testing dataset. Our matched-filtering approach, which we used with a single exact template, provides an unrealistic upper-bound to the sensitivity since a full template-bank analysis would suffer from a higher false alarm rate. The \ac{ROC} curve for the \ac{LSTM} model, still in early development, shows that at fiducial false alarm rates of 1 per day, 1 per 6 hours, and 1 per hour has a true alarm probability $\sim 2$ times lower than the upper-bound. We note that a rigorous matched-filter analysis using a complete template bank would likely yield results somewhere between the 2 \ac{ROC} curves shown. The fact that the \ac{LSTM} result is within a factor of 2 in sesnitivity of a best case scenario is important because the \ac{LSTM} model could be improved further and offers faster, more efficient processing. 

\subsection{Early Warning times from the LSTM Model} \label{subsec: 5.2}

\begin{figure}
\centering
{\includegraphics[width=0.55\textwidth]{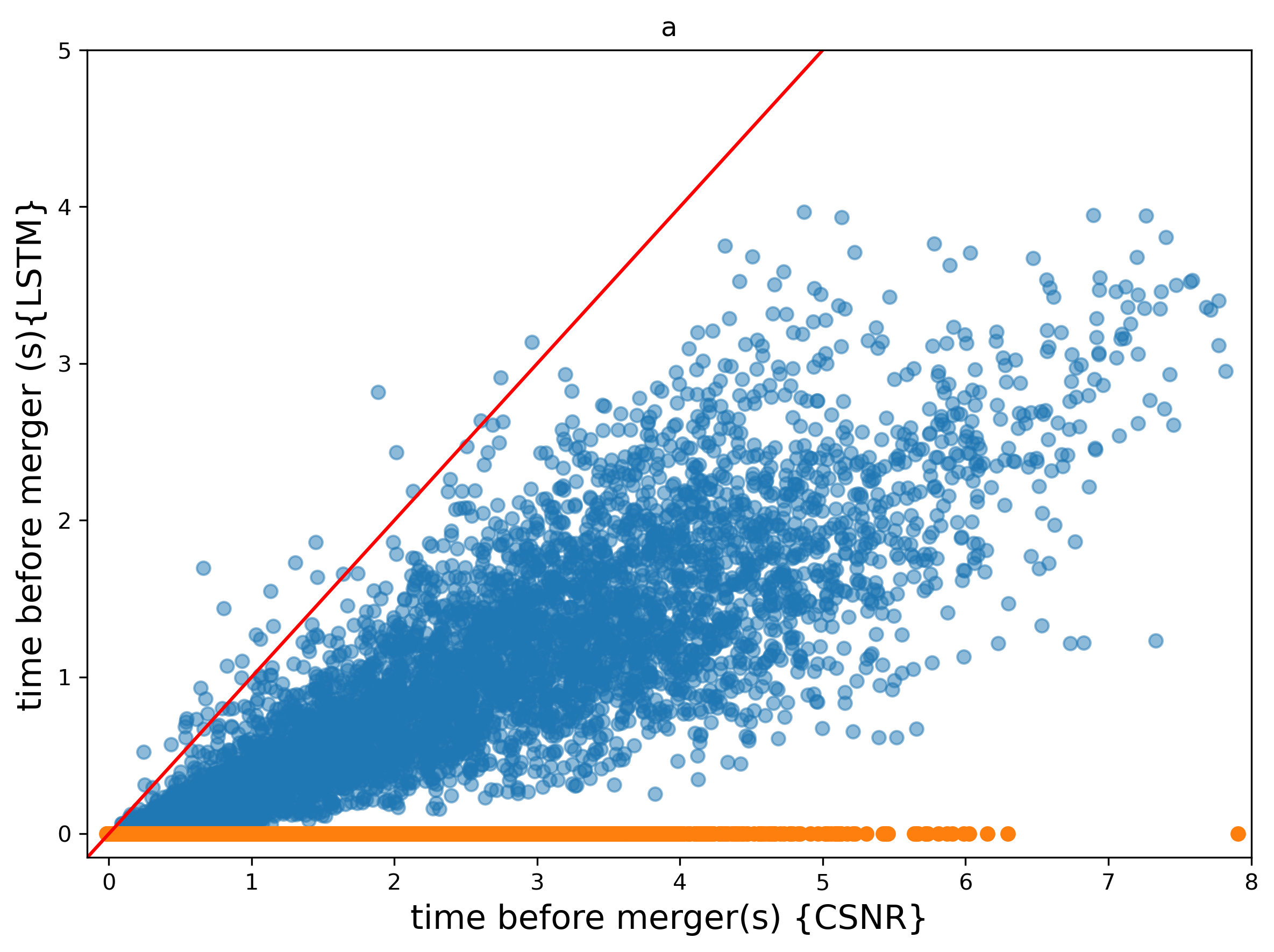}}
{\includegraphics[width=0.55\textwidth]{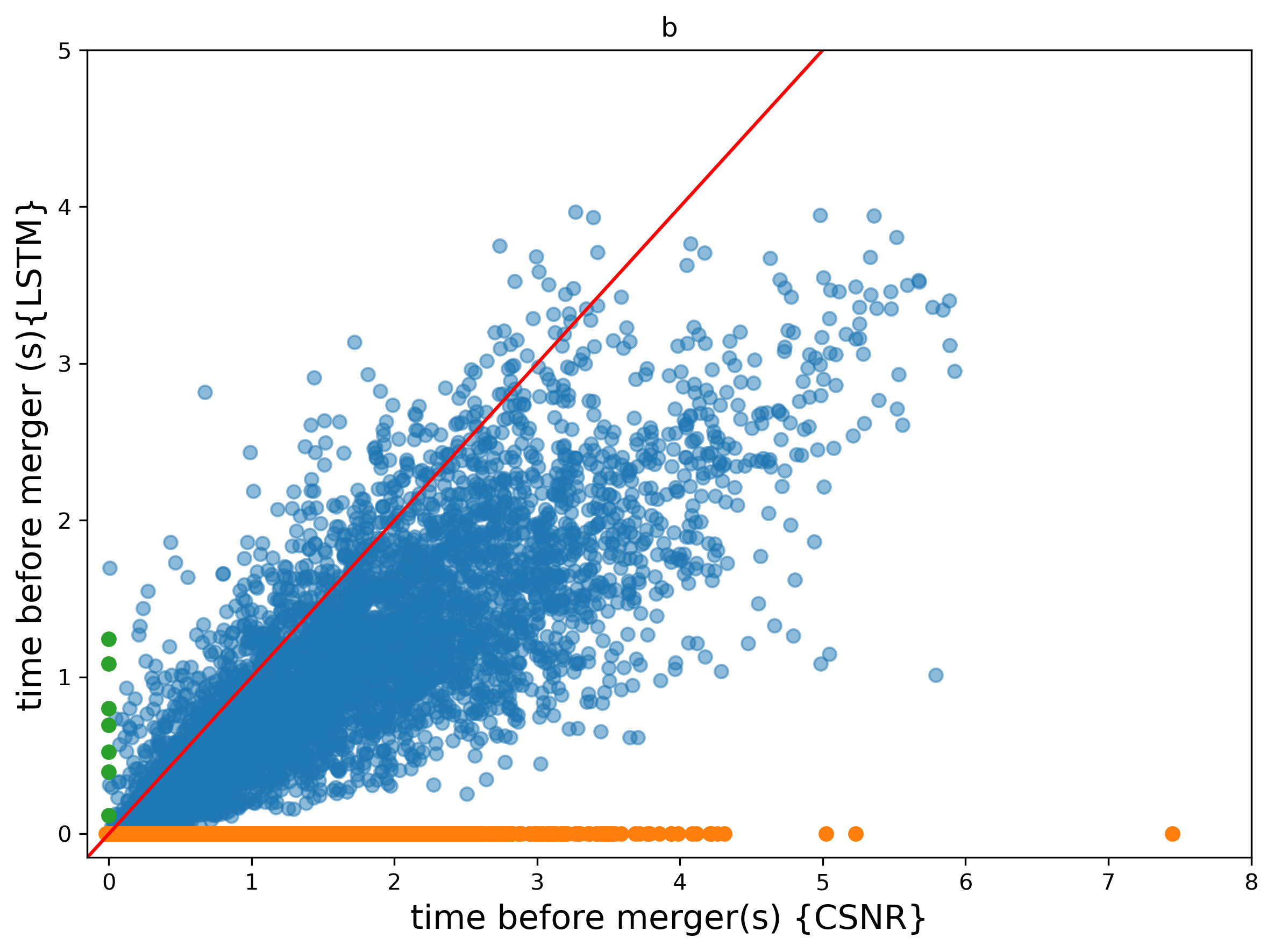}}
{\includegraphics[width=0.55\textwidth]{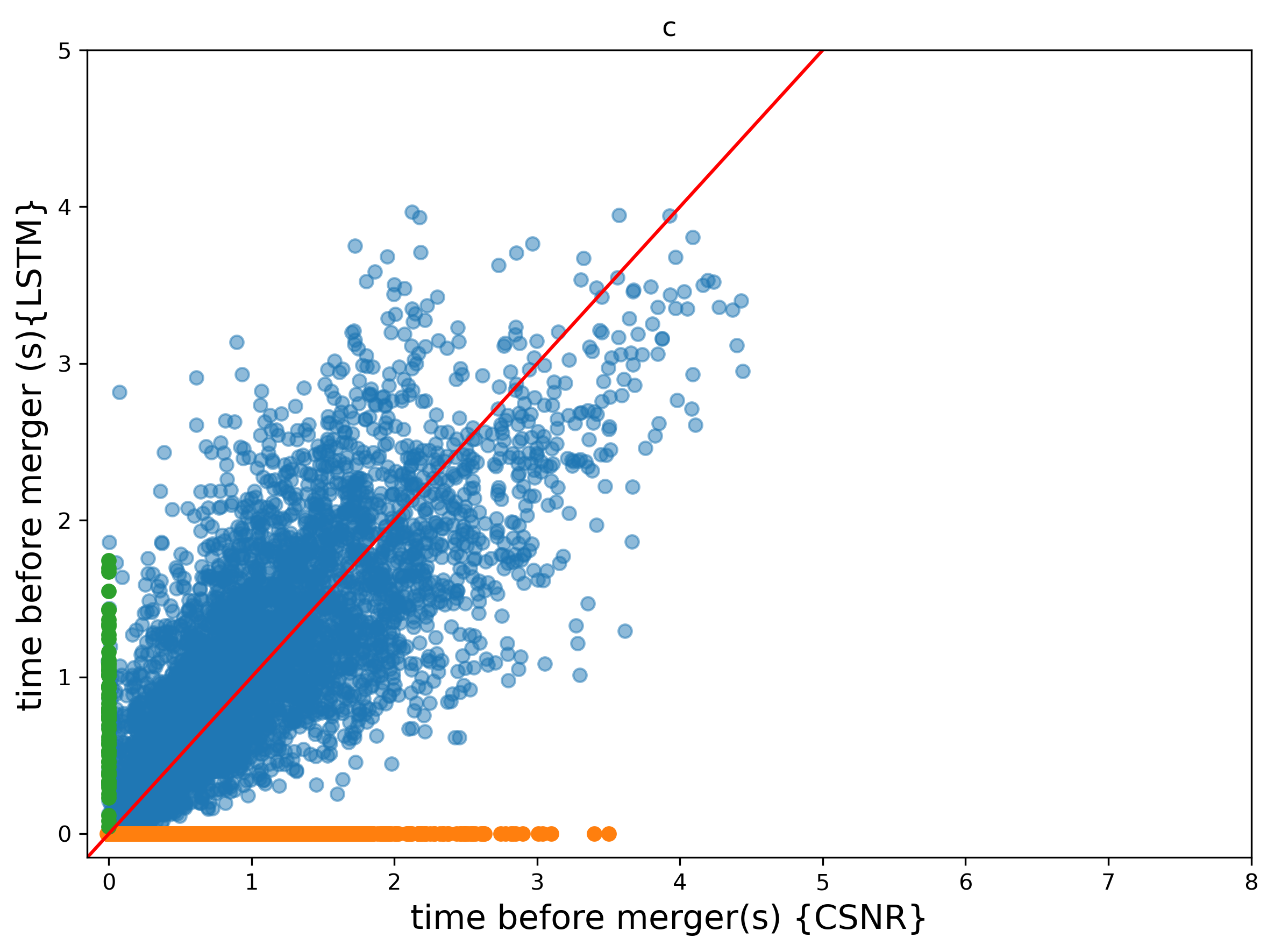}}
\caption{Early Warning Detection Analysis at a fixed false alarm rate of one per day: Plots (a), (b), and (c) display the performance of the \ac{LSTM} model and cumulative \ac{SNR} thresholds, set at 6, 8, and 10 respectively, for early warning detection. Green dots signify early warnings detected solely by the \ac{LSTM} model, not identified by cumulative \ac{SNR}. Orange dots indicate early warnings detected only by cumulative \ac{SNR}. Blue dots represent common detections by the \ac{LSTM} model and cumulative \ac{SNR}.} 
\label{fig: 5}
\end{figure}

Here we provide the results of our early warning analysis, which focuses on the \ac{LSTM} model's and \ac{CSNR}'s comparative performance in the context of early warning detection. As a proxy for comparing both analyses at common false alarm rates we instead provide comparative results assuming 3 different \ac{CSNR} thresholds (6,8, and 10) for the approximate matched-filter benchmark. In Fig.~\ref{fig: 5} we see a significant positive correlation between the early warning times calculated by the approximation \ac{CSNR} technique and those determined by the \ac{LSTM} model. Most importantly, it is found that, in terms of early warning capabilities, a \ac{CSNR} threshold of 10 corresponds to a performance comparable to the \ac{LSTM}. Moreover, it is noteworthy that the early warning time provided by the \ac{LSTM} is around half that provided by the \ac{CSNR} approach at a \ac{CSNR} threshold of 6 and reaches roughly two-thirds at a threshold of 8. 
\begin{table}
\centering
\caption{Early Warning Detection Analysis of the \ac{LSTM} Compared with Detection of the \ac{CSNR} at Different Thresholds (6, 8, and 10)}
\begin{tabular}{|c|c|c|c|}
\hline \hline
Threshold Value \ac{CSNR} & 6 & 8 & 10  \\
\hline
 Early warning detected by \ac{LSTM} and \ac{CSNR} & 38.52\% & 44.71\% & 52.57\%    \\
Early warning detected by \ac{LSTM} only   & 0.0\% & 0.06\% &  0.66\%    \\
Early warning detected by \ac{CSNR} only  &  61.48\% & 55.16\%  &  46.70\%  \\

\hline \hline
\end{tabular}
\label{tab: 2}
\end{table}

\begin{figure}
\includegraphics[width= 1 \linewidth]{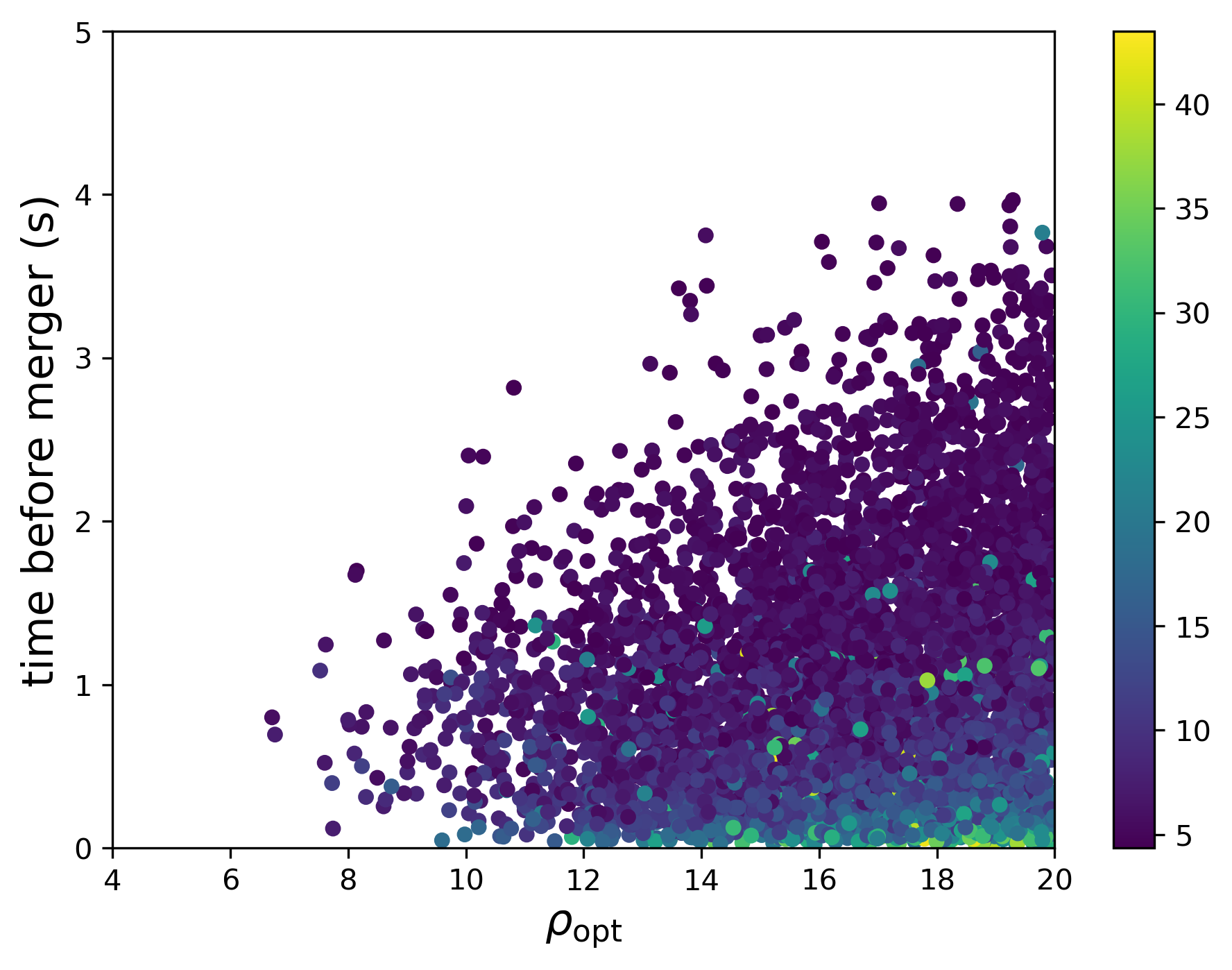}
\centering
\caption{Illustrating the \ac{LSTM} model's capability for early event warnings based on the signal, as a function of optimal \ac{SNR}, this graph spans \ac{SNR} values ranging from 2 to 20. The $x$-axis represents the \ac{SNR}, while the $y$-axis denotes the amount of advance time the \ac{LSTM} model can predict an event before it occurs. Additionally, the associated colors in the plot represent the chirp mass of the signals, providing a visual correlation between the mass and the model's predictive timing. These results correspond to a false alarm rate of one per day. 
\label{fig: 6}}
\end{figure}

The \ac{LSTM} model can produce early warning up to 4 seconds before merger, depending on the \ac{SNR} and the chirp mass of the system. For signals with an \ac{SNR} greater than 5, the \ac{LSTM} successfully detects some signals and provides an early warning alert. This early warning is generated based on a specific threshold parameter selected according to a predetermined false alarm rate. Fig. \ref{fig: 6} analyzing The efficacy of the \ac{LSTM} model in providing early warnings of signals is analyzed with respect to the \ac{SNR}. It is evident that higher \ac{SNR} values, when associated with lower chirp masses, significantly increase the model's early warning detection capabilities. Conversely, at lower \ac{SNR} levels, the model's ability to provide early warnings appears to diminish, suggesting a direct correlation between \ac{SNR} and the effectiveness of early warning signals. Additionally, the \ac{SNR} plays a critical role; higher \ac{SNR} values are associated with increased lead times, enhancing the model's early warning capabilities.

\begin{figure}
\includegraphics[width= 1 \linewidth]{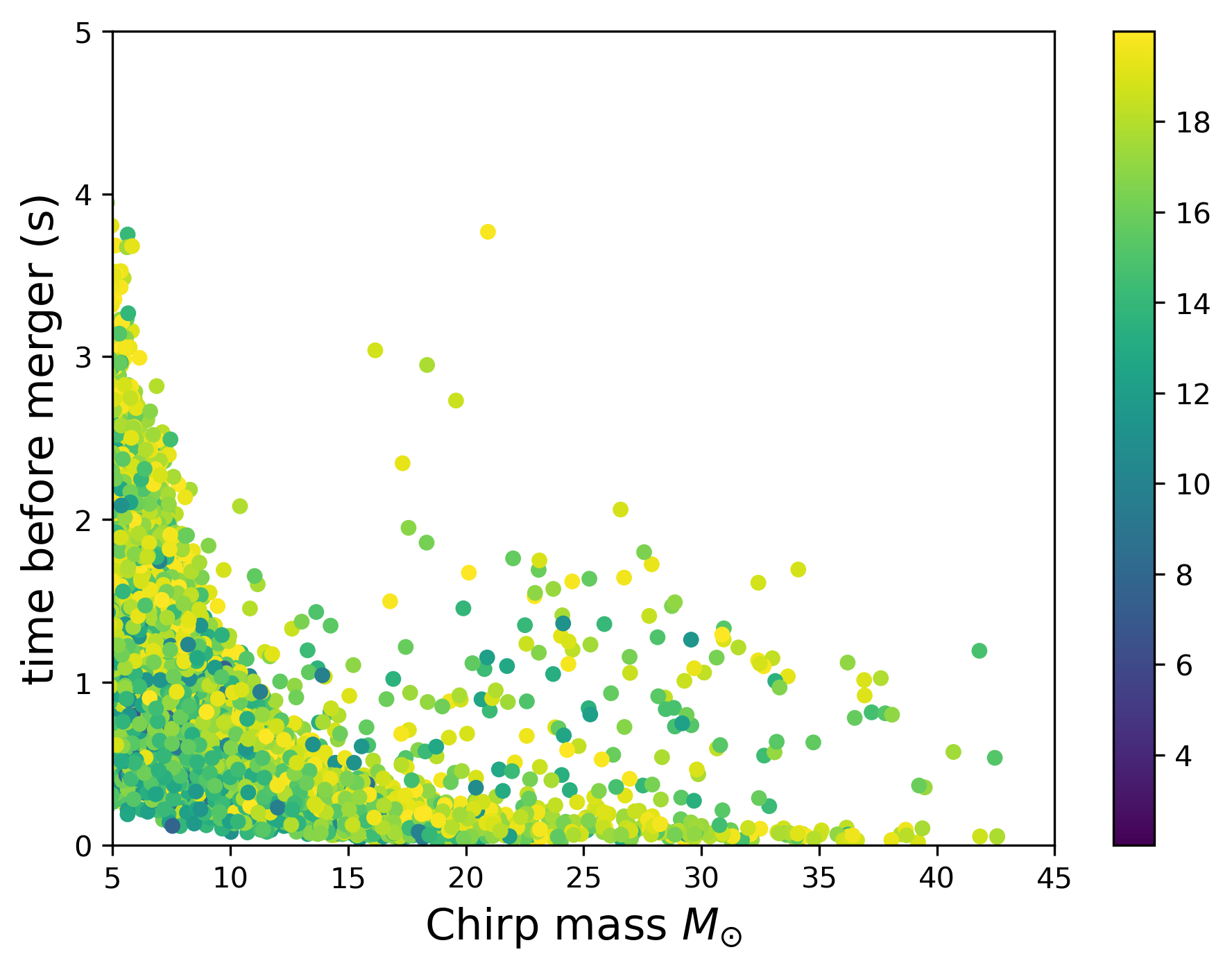}
\centering
\caption{The LSTM model's capacity for early warning detection is shown as a function of chirp mass on the $x$-axis, while the $y$-axis displays the model's lead time in predicting events before they happen. The color gradient represents the \ac{SNR} , providing an understanding of the relationship between \ac{SNR} levels and the model's advance prediction time. These results are based on a false alarm rate of one per day. 
\label{fig: 7}}
\end{figure}

The scatter plot in Fig.~\ref{fig: 7} demonstrates the LSTM model's predictive performance in early warning detection as a function of chirp mass. A clear trend is observed where lower chirp mass values, coupled with higher \ac{SNR}, result in a greater lead time for early warning detection. Conversely, as chirp mass increases, the likelihood of obtaining an early warning diminishes, verifying the expected inverse relationship between chirp mass and the efficiency of early warning detection. We expect lower chirp mass systems to be longer in duration and hence provide the opportunity for early detection.  

\begin{figure}
\includegraphics[width= 1 \linewidth]{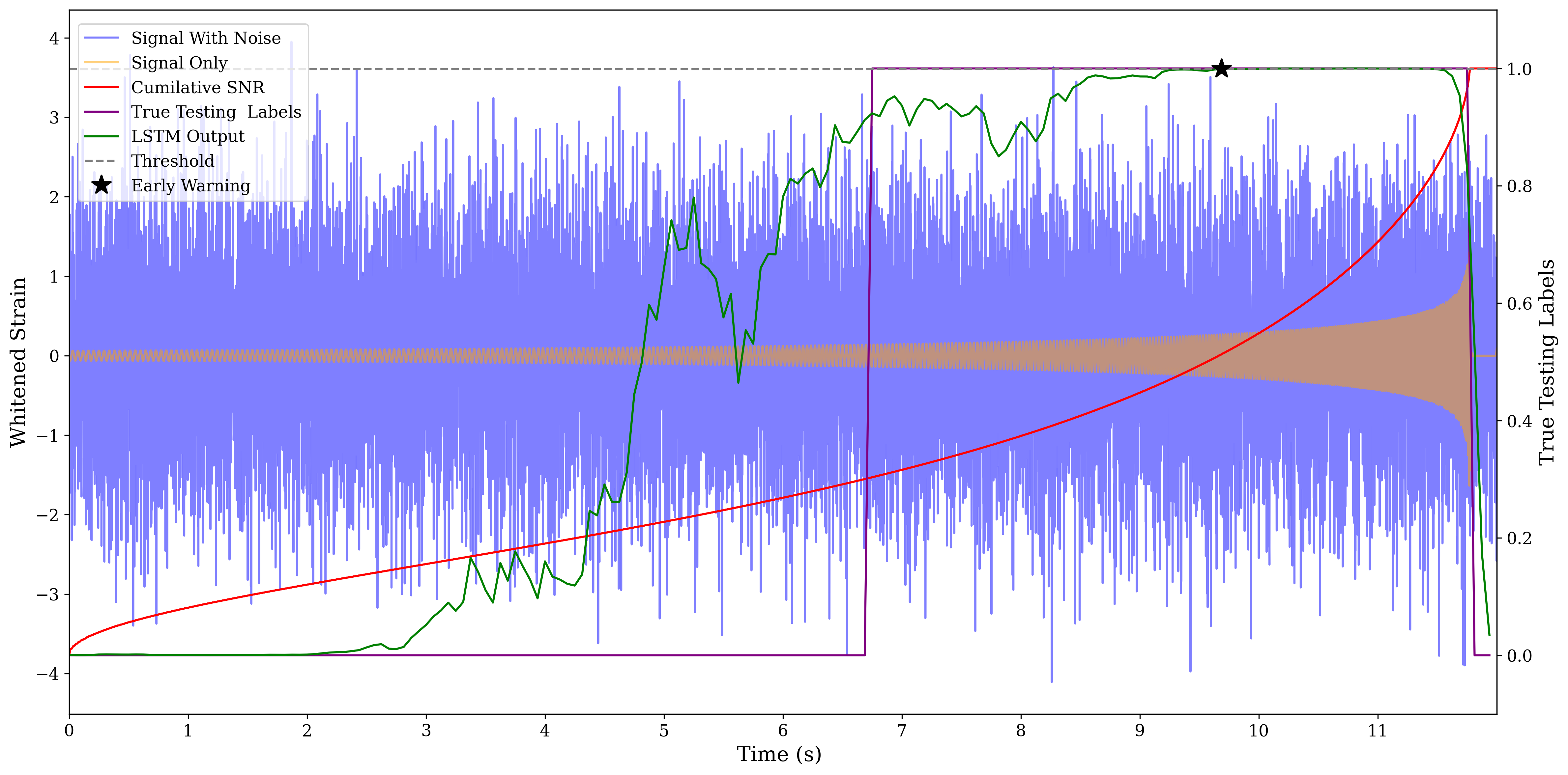}
\centering
\caption{An example output of the \ac{LSTM} model for a \ac{BBH} testing signal in Gaussian noise that merged within the observation window. \ac{LSTM} Output which is represented by (green line) that can detect the signal at a FAR of $1$ per day, approximately $2$ seconds before it has merged, where the orange line represents the signal and the blue line represent the waveform with noise. The red line represents the cumulative \ac{SNR}, and the purple line represents the True labels. \label{fig: 8}}
\end{figure}

\begin{figure}
\includegraphics[width= 1 \linewidth]{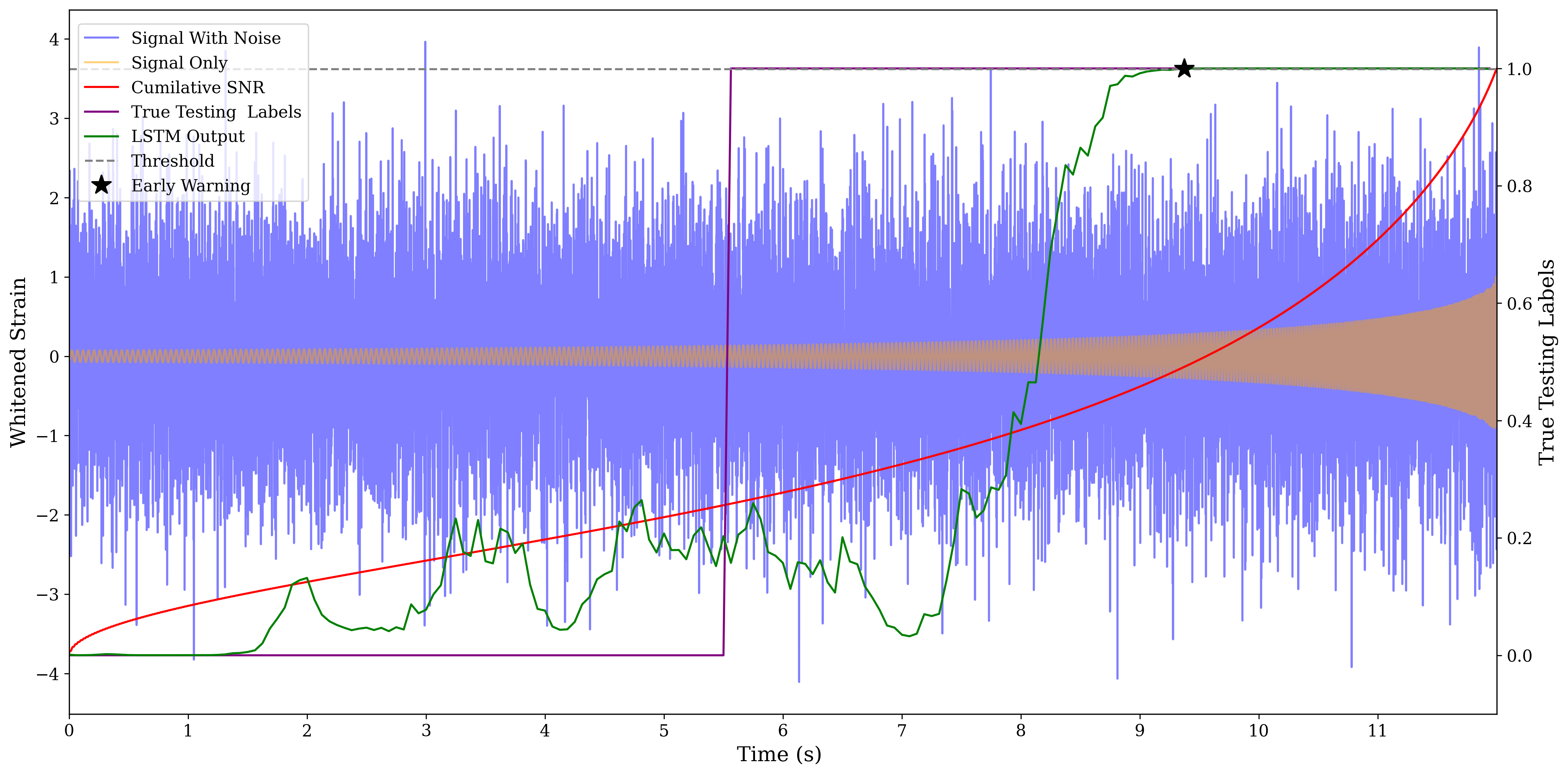}
\centering
\caption{An example output of the \ac{LSTM} model for a \ac{BBH} testing signal in Gaussian noise, which is represented by signal statistics (green line) that can detect the signal  $\sim 2$ seconds before it has merged, where the orange line represents the signal and the blue line represent the waveform with noise. The red line represents the cumulative \ac{SNR}, and the purple line represents the labels at a FAR of 1 per day. 
\label{fig: 9}}
\end{figure}

\begin{figure}
\includegraphics[width= 1 \linewidth]{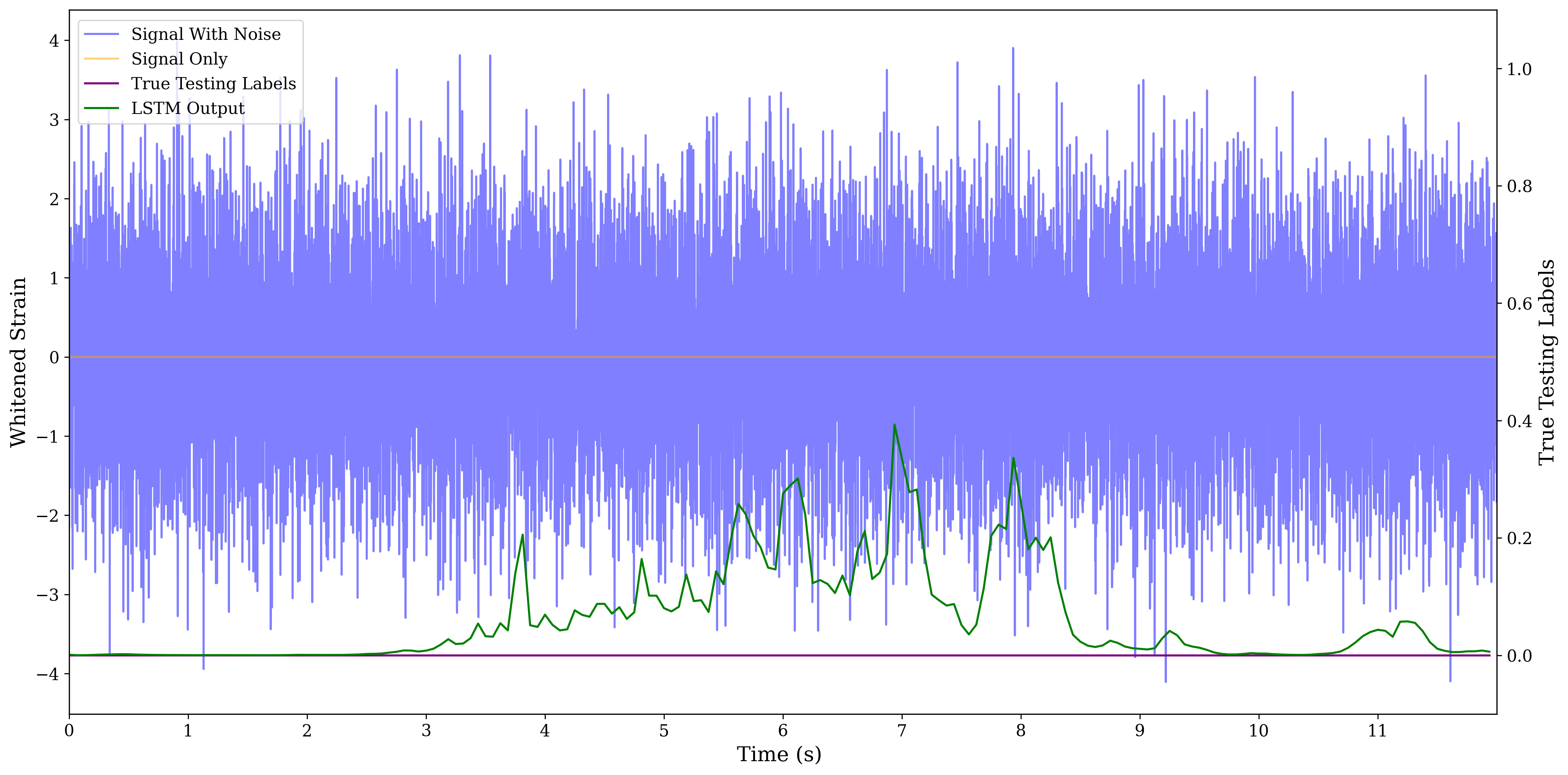}
\centering
\caption{Another example output of the \ac{LSTM} model, this time detection statistic provided results consistent with there being no signal present in the data at a FAR of 1 per day, as indicated by the signal statistics (green line) approaching zero. The orange line represents the signal, the blue line depicts the signal with noise, and the purple line denotes the labels.
\label{fig: 10}}
\end{figure}

%
Figures~\ref{fig: 8}, \ref{fig: 9}, and \ref{fig: 10} illustrate the performance of the \ac{LSTM} model when applied to different test data time-series. Specifically, Fig. \ref{fig: 8} demonstrates the model's capability to detect the signal prior to merger. The LSTM model successfully identifies the signal approximately 2 seconds before the merger which is represented by the $\star$ where the  green line crosses the false alarm threshold. The characteristics of this detected signal include a \ac{SNR} of 16.52 and a chirp mass of 4.81 $M_{\odot}$, merging at 11.77 seconds within the time window under consideration.

In a separate instance shown in Fig.~\ref{fig: 9}, the model detects another signal roughly 3 seconds prior to its merger. This signal is distinguished by a \ac{SNR} of 19.5 and a chirp mass of 5.17 $M_{\odot}$. In this case, the merger occurs at 12.15 seconds, falling outside the time window analyzed. These examples underscore the LSTM model's efficacy in temporal signal detection within varied observational windows.

Figure. \ref{fig: 10} presents the performance of the \ac{LSTM} model in recognizing the absence of the signal, where only noise is present. It is evident from the \ac{LSTM} output that, in the absence of a signal, the model's predictions approach zero, indicating its effectiveness in differentiating between noise and actual signals. This capability is crucial for reducing false positives in signal detection and enhances the reliability of the model in practical applications.

\begin{figure}
\includegraphics[width= 1 \linewidth]{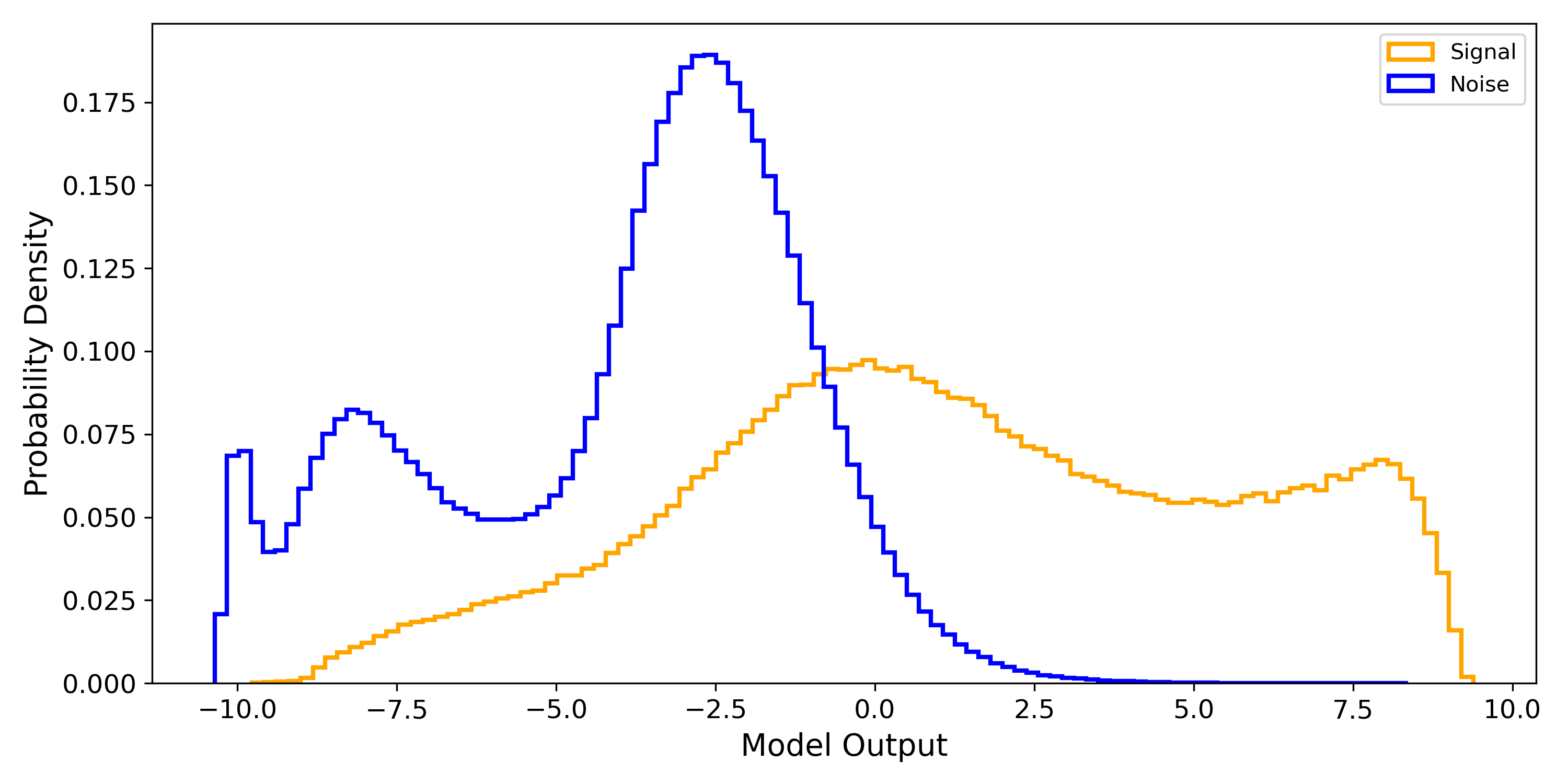}
\centering
\caption{Distribution of the \ac{LSTM} model's classifications of signal and noise based on test data evaluation. The blue curve represents the probability density of instances classified as noise, whereas the orange curve shows the distribution of instances classified as signal by the \ac{LSTM} model
\label{fig: 11}}
\end{figure}

Figure~\ref{fig: 11} presents the probability distribution of \ac{LSTM} output (prior to the \ac{BCE} sigmoid activation to convert to a probability) with test data as input. The signal distribution exhibits a rightward shift, indicating a higher probability of correct signal detection. In contrast, the noise distribution is shifted leftward, reflecting the model's effectiveness in separating noise from signal. The overlap in the central region suggests the presence of signals with low \ac{SNR}, where the model's distinction between signal and noise becomes less pronounced and also indicates a subset of signals that are challenging to classify. These are instances where the signals are relatively weak and, hence, more susceptible to misclassification as noise. Conversely, the tail on the far right signifies signals with high \ac{SNR}, which the \ac{LSTM} model has identified with a high degree of confidence. These central cases may represent signals partially obscured by noise at the beginning of the signal time series. Yet, such events will yield robust detections where the signal's strength is pronounced enough to be clearly discernible from the noise. 

\section{Conclusion and Discussion} \label{sec: 6}

In this study, we provide a novel \ac{LSTM}-based early warning system for \ac{BBH}s signals. We have demonstrated that \ac{BBH}s events can be identified even in cases where the data only contain a portion of the early inspiral. The methodology we have described has the potential to be extended and applied effectively to detect various types of mergers, including those involving \ac{BNS}s and \ac{NSBH} systems. In summary, using the \ac{LSTM} networks in \ac{GW}s early warning detection from \ac{BBH}s signals has shown encouraging results. Early detection of \ac{GW}s signals is a crucial prerequisite for time-sensitive astronomical observations, and the \ac{LSTM} model has demonstrated its ability to achieve this. Moreover, an evaluation related to conventional matched filtering methods indicates that the sensitivity of \ac{LSTM} networks is competitive, providing a workable substitute with certain advantages. One of the most significant benefits of using \ac{LSTM}s is their efficiency in terms of computational resources. Unlike more complex models that may require substantial GPU memory, \ac{LSTM}s are lightweight and can be run with relatively low memory overhead. This makes them particularly suitable for applications where low latency is crucial. With the ability to operate effectively in low-latency scenarios, \ac{LSTM} networks can facilitate rapid data processing, which is essential for prompt scientific response and follow-up observations. 

Additionally, the \ac{LSTM} model also performs exceptionally well regarding flexibility to various data ranges. Once trained on a dataset, the model is fast to test over different data ranges and exhibits good generalization capabilities. This flexibility makes it possible to quickly adjust to new data without requiring thorough retraining. This is very helpful in the dynamic field of \ac{GW}s astronomy, where the swift detection and reporting of events can have significant scientific ramifications. For instance, in the context of multi-messenger follow-ups, the ability of \ac{LSTM}s to identify potential \ac{GW}s signals promptly can enable astronomers to coordinate observational resources across the \ac{EM} spectrum and other cosmic messengers, such as neutrinos or cosmic rays. This coordinated approach can lead to a more comprehensive understanding of the events, further cementing the role of \ac{LSTM} models as a potentially valuable asset in the era of time-domain astronomy.

The \ac{LSTM} model, with its competitive sensitivity and flexibility, offers us a new tool for the early detection of \ac{GW}s. Moreover, this model is especially promising for systems like \ac{BNS}s, where there is a chance to detect \ac{EM} or neutrino counterparts and a significantly longer early warning period. In our current research, certain aspects were excluded, which we acknowledge as avenues for enhancement in future investigations. These elements include the integration of multiple detectors, the consideration of spinning systems, and the incorporation of real detector noise. Moreover, incorporating real detector noise is critical for improving the model's practical applicability and resilience in real-world scenarios. Future work could readily adapt our existing framework to include these elements.





\section*{Acknowledgements}

The authors thank Siong Heng and other members of the Data Analysis Group of the Institute of Gravitational Research for helpful discussions. C.M. is supported by the Science and Technology Research Council [ST/V005634/1]. This material is based upon work supported by NSF's LIGO Laboratory which is a major facility fully funded by the National Science Foundation.

\renewcommand{\bibname}{References}
\bibliographystyle{abbrv}
\bibliography{References}


\end{document}